\documentclass[aps,prb,10pt,twocolumn,superscriptaddress]{revtex4-2}

\usepackage{graphicx}
\usepackage{graphics}
\usepackage{amsmath}
\usepackage{amssymb}
\usepackage{amsfonts}
\usepackage{dsfont}
\usepackage{braket}
\usepackage{color}
\usepackage{braket,slashed}
\usepackage[mathscr]{euscript}
\definecolor{darkblue}{rgb}{0, 0, 0.8}
\usepackage[colorlinks=true, breaklinks=true, linkcolor=red, citecolor=blue, urlcolor=blue]{hyperref} 
\usepackage{hyperref}
\usepackage{subfigure}
\usepackage{xfrac}
\usepackage{bm}
\usepackage{kantlipsum}
\usepackage{enumitem}
\usepackage{tikz}
\usepackage{framed}
\usepackage{graphicx}
\usepackage{subfigure}
\usepackage{cleveref}
\usepackage{array}

\newcommand{\parL}[1]{\noindent\textbf{\textit{#1}}---}

\allowdisplaybreaks[1]

\newcommand{\textmath}[1]{\ensuremath{\text{\textit{#1}}}}
\newcommand{\mathtext}[1]{\ensuremath{\text{\textit{#1}}}}
\newcommand{\mt}[1]{\ensuremath{\text{\textit{#1}}}}

\newcommand{\vect}[1]{\ensuremath{\boldsymbol{#1}}}
\renewcommand{\dag}{^\dagger}
\newcommand{\pdag}{^{\phantom{\dagger}}}
\renewcommand{\vr}{\ensuremath{\vect{r}}}
\newcommand{\vv}{\ensuremath{\vect{v}}}
\newcommand{\vA}{\ensuremath{\vect{A}}}
\newcommand{\vB}{\ensuremath{\vect{B}}}

\newcommand{\mT}{\ensuremath{\mathcal{T}}}
\newcommand{\e}{\ensuremath{e}}

\newcommand{\U}{\ensuremath{\mathrm{U}}}

\newcommand{\one}{\ensuremath{\mathds{1}}}

\newcommand{\hc}{\ensuremath{\textmath{h.c.}}}

\begin{document}

\title[Gauge-covariant projected entangled paired states for interacting systems in a magnetic field]{Gauge-covariant projected entangled paired states\\for interacting systems in a magnetic field}

\author{Wei Tang}
\email{wei.tang.phys@gmail.com}
\affiliation{Department of Physics and Astronomy, Ghent University, Krijgslaan 281, 9000 Gent, Belgium}

\author{Gunnar M\"oller}
\affiliation{Physics of Quantum \& Materials Group, School of Engineering,
Mathematics and Physics, University of Kent, Canterbury CT2 7NH, United Kingdom}

\author{Frank Verstraete}
\affiliation{Department of Physics and Astronomy, Ghent University, Krijgslaan 281, 9000 Gent, Belgium}
\affiliation{Department of Applied Mathematics and Theoretical Physics, University of Cambridge, Wilberforce Road, Cambridge, CB3 0WA, United Kingdom}

\author{Laurens Vanderstraeten}
\email{laurens.vanderstraeten@ulb.be}
\affiliation{Center for Nonlinear Phenomena and Complex Systems, Université Libre de Bruxelles, Belgium}

\date{\today}

\begin{abstract}
The Hamiltonian for a system of itinerant particles on a two-dimensional lattice in a uniform magnetic field reduces the translational symmetry to a magnetic translation group, because of the need to choose a particular gauge for the vector potential. Nonetheless, in many situations all physical observables of the ground state remain entirely translation invariant. In this work, we introduce a projected entangled-pair state (PEPS) wavefunction with a pattern of virtual flux tensors, for which all physical expectation values are translation invariant by construction, possibly within an enlarged unit cell reflecting any symmetry breaking in the target state. Moreover, we show that the usual contraction and optimization methods for translation-invariant PEPS can be used, with the magnetic flux per plaquette only entering as a continuous parameter in the tensor network contractions. Therefore, our approach provides a method for simulating an interacting many-body system in a uniform magnetic field independently of the gauge choice for the vector potential and bypassing the need to consider extended magnetic unit cells.
\end{abstract}

\pdfstringdefDisableCommands{\def\\{ }}
\maketitle

\newcommand{\diagram}[1]{\vcenter{\hbox{\includegraphics[scale=0.65,page=#1]{./diagrams_main.pdf}}}}

\parL{Introduction}%
%
In quantum mechanics, the motion of a particle with mass $m$ and charge $q$ in a uniform magnetic field $\vB$ is described by the Hamiltonian
\begin{equation}
    H = \frac{1}{2m} \bigl[ - i \vect{\nabla} - q \vA(\vr) \bigr] ^2 \; ,
\end{equation}
where $\vA$ is the vector potential corresponding to the magnetic field, i.e. $\vB = \vect{\nabla}\times\vA$. There is a freedom of gauge choice for the vector potential which entails a corresponding transformation of the single-particle wavefunction, following
\begin{align} 
    \vA(\vr) &\to \vA(\vr) + \vect{\nabla} \chi(\vr) \;  \label{eq:gauge_trf} \\
    \Psi(\mathbf{r}) &\to e^{i \chi(\mathbf{r})} \Psi(\mathbf{r}), \label{eq:gauge_trf_state}
\end{align}
implying that the Hamiltonian transforms covariantly with $\vA$. Although classical equations of motion can always be formulated without an explicit dependence on the vector potential, it necessarily appears in the quantum-mechanical description of a charged particle. 

\begin{figure}[t]
    \centering
    \includegraphics[width=0.7\columnwidth,page=1]{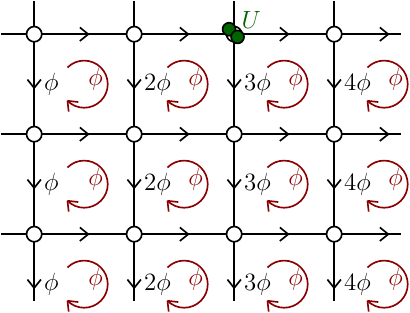}
    \caption{Graphical representation of the Harper-Hofstadter-Hubbard Hamiltonian [Eq.~\eqref{eq:hof_hub}] with hopping strength $t$, on-site repulsion $U$ and magnetic flux per plaquette $\phi=2\pi a^2B/\Phi_0$. We have chosen the Landau gauge $\vA=-Bx\, \vect{e}_y$, leading to the indicated pattern of Peierls phases $\phi_{ij}$ in the hopping terms $-t (e^{i\phi_{ij}} b_i^\dagger b_j + h.c.)$. The arrows indicate the orientation of the line integral defining $\phi_{ij}$, while the labels $\phi, 2\phi, 3\phi,\ldots$ specify the magnitude of the phase accumulated along that direction.}
    \label{fig:hofstadter}
\end{figure}

This situation has particularly interesting consequences in two-dimensional condensed-matter systems with an orthogonal uniform magnetic field, chosen here as $\vect{B}=-B\,\vect{e}_z$. In lattice systems, the magnetic field is typically incorporated by writing down a tight-binding model where the hopping terms acquire a phase factor \cite{Luttinger1951, Kohn1959}. This Peierls substitution \cite{Peierls1933} leads to the following form for the Hamiltonian in second quantization,
\begin{align}
    H_{\mathtext{HH}} &= - \, \sum_{\braket{ij}} e^{i\phi_{ij}} b_i\dag b_j\pdag + e^{-i\phi_{ij}} b_j\dag b_i\pdag \nonumber \\ &\hspace{2cm}\text{with}\quad \phi_{ij} = \frac{2\pi}{\Phi_0} \int_{\vr_j}^{\vr_i} \vA(\vr) \cdot d\vr
\end{align}
often called the Harper-Hofstadter model \cite{Harper1955, Azbel1964, Hofstadter1976}. Here the $(b_i\dag,b_i\pdag)$ are the usual creation and annihilation operators at site $i$ in the lattice, and $\Phi_0$ is the elementary magnetic flux. On the square lattice, one particularly convenient gauge choice is the Landau gauge $\vA=-Bx\, \vect{e}_y$, which gives rise to a pattern of phases in the hopping terms as the one shown in Fig.~\ref{fig:hofstadter}. Physically, however, the only relevant parameter is the flux per plaquette $\phi=2\pi a^2 B/\Phi_0$, with $a$ the lattice spacing.

In the many-body setting of particles on a lattice, a gauge transformation parametrized by $\chi(\vr)$ [Eq.~\eqref{eq:gauge_trf}] is a local diagonal unitary transformation of the form
\begin{equation}
U({\chi(\vr)}) = \prod_j \; U_j\left(\frac{2\pi i}{\Phi_0} \chi(\vr_j)\right),
\label{eq:gauge_trf_U}
\end{equation}
where $U_j(\phi)=\e^{i\phi n_j}$ is a local $\U(1)$ group action at site $j$ with $n_j=b_j\dag b_j\pdag$ the local number operator. Any particular choice for the vector potential necessarily reduces the translation symmetry of the Hamiltonian. Instead, the Hamiltonian is invariant under the combination of translation symmetry and a gauge transformation, leading to a generalized translation operator over the lattice vector $\vv$, 
\begin{equation} \label{eq:Tv}
    \tilde{\mT}_{\vv} = U({\chi(\vr)}) \mT_{\vv}\;,
\end{equation} 
with the gauge transformation $\chi(\vr)$ chosen such that it undoes the translation of the vector potential. These operators no longer commute for translations in different lattice directions, and instead give rise to a magnetic translation group \cite{Brown1964, Zak1964}. For a rational value of the flux $\phi=2\pi p/q$ with $p$ and $q$ co-prime, the magnetic unit cell encompasses $q$ lattice sites or plaquettes.

These symmetry considerations remain valid in the interacting many-body case. For bosons, a minimal model is provided by the Harper-Hofstadter-Hubbard Hamiltonian
\begin{equation} \label{eq:hof_hub}
    H_{\mathtext{BHHH}} = t\,H_{\mt{HH}} + \frac{U}{2} \sum_i n_i (n_i-1),
\end{equation}
where the parameters are the hopping strength $t$, the interaction strength $U$, the flux per plaquette $\phi$ and the filling $\rho$. Although analytical treatments for this model are possible in special limiting cases, one typically needs to perform numerical simulations in order to get a quantitatively accurate picture of the phase diagram. Exact diagonalization \cite{Sorensen2005, Moller2009, Sterdyniak2012, Laeuchli2013, Moeller2015} is severely limited by the need to choose system sizes that are commensurate with the magnetic unit cell, and quantum Monte Carlo methods exhibit sign problems due to the phases in the hopping terms. Alternatively, tensor network \cite{Cirac2021, Xiang2023} methods can be used on finite geometries with tree tensor networks \cite{Gerster2017} and on elongated cylinders using matrix product states (MPS) \cite{Kovrizhin2010, Zhu2015, Grushin2015, Motruk2015}. More recently, projected entangled-pair states \cite{Verstraete2004} (PEPS) were used directly in the two-dimensional thermodynamic limit for the bosonic \cite{Weerda2024} and the fermionic \cite{Niu2026, Chen2026} versions of this model, but the need for accommodating the magnetic unit cell puts a stringent limitation of the values for the flux.

Physically, however, we expect that implementing these large unit cells is overkill: While the Peierls substitution introduces position-dependent phases in the hopping amplitudes, this does not imply a physical breaking of translation symmetry. Rather, the system remains invariant under the magnetic translation group, in which lattice translations are accompanied by appropriate gauge transformations. In particular, expectation values of local observables may remain translation invariant even when the microscopic representation of the Hamiltonian is not. 

In this work, we introduce a PEPS ansatz that is explicitly invariant under these magnetic translations, in the sense explained below. In this way, the variational parameters of the ansatz are cleanly separated conceptually from the choice of gauge, and may reduce to a single translationally invariant tensor whenever the underlying phase does not spontaneously break translation symmetry. In addition, we develop a PEPS contraction algorithm that respects this translation symmetry explicitly, which allows to optimize the variational energy as a function of a single PEPS tensor without breaking any symmetries. We provide benchmark simulations that showcase the performance and utility of this approach.

\parL{PEPS ansatz}%
%
Due to the extensivity property of tensor network states, a translation-invariant PEPS wavefunction \cite{Verstraete2004} can be formulated directly in the thermodynamic limit by repeating a single local tensor $A$, represented diagrammatically as
\begin{equation}
\diagram{18} \;.
\end{equation}
The $\U(1)$ charge conservation symmetry of the model can be explicitly encoded as a local $\U(1)$ symmetry of each PEPS tensor separately, where the charge density of the state is encoded by adding an auxiliary leg with a single non-trivial charge (this leg is denoted by a wavy line in the diagram). The $\U(1)$ symmetry of the local PEPS tensor requires that the virtual spaces also have a $\U(1)$ charge structure $\{q_v\}$ (possibly with additional degeneracies), and that the physical action of a group element can be dragged to the virtual legs:
\begin{equation} \label{eq:u1_peps}
    \diagram{3} = \diagram{4} \;.
\end{equation}
where the group elements of the $\U(1)$ symmetry are denoted as
\begin{equation}
    \diagram{5} = \delta_{p,q}\e^{i q \phi} \one_{D_p} \;.
\end{equation}
Using these local $\U(1)$ symmetric $A$ tensors, an extensive PEPS wavefunction is now written as \cite{Tang2025}
\begin{equation}
    \ket{\Phi(A)} = \diagram{2} \;,
\end{equation}
where we have introduced bond tensors without any variational parameters 
\begin{equation} \label{eq:round}
    \diagram{6} = \delta_{p,-q} \one_{D_p} \;.
\end{equation}
Depending on the physical situation, in addition to the internal $\U(1)$ symmetry we can also impose spatial symmetries such as rotation and reflection, and restrict to real tensor entries.

We now incorporate a magnetic field by first realizing that a phase-modulated hopping term on two sites can be written in terms of a unitary transformation with the local $\U(1)$ group actions :
\begin{multline}
    \e^{i\phi_{12}} b_1\dag b_2\pdag + \e^{-i\phi_{12}} b_2\dag b_1\pdag \\ = U_1(\phi_1)\dag U_2(\phi_2)\dag \left( b_1\dag b_2\pdag + b_2\dag b_1\pdag \right) U_1(\phi_1)U_2(\phi_2),
\end{multline}
where we parametrize the Peierls phase $\phi_{12}$ locally as a difference of site-dependent phases, $\phi_{12} = \phi_1 - \phi_2$.
To capture their effect in the PEPS wavefunction, we consider the action of these local unitary representations on two consecutive PEPS tensors: due to the $\U(1)$ symmetry [Eq.~\eqref{eq:u1_peps}], these group actions can be dragged to the virtual level to obtain
\begin{multline}
    \diagram{7} \\ = \diagram{8} \;,
\end{multline}
i.e., we obtain a virtual group action with the Peierls phase $\phi_{12}$ on the bond linking the two tensors. This implies that we can implement a phase modulation of a hopping term locally in the PEPS wavefunction by placing a $\U(1)$ group action on the virtual level. This operation is now repeated on all links: we imprint a pattern of Peierls phases into the PEPS wavefunction by adding a corresponding pattern of virtual $\U(1)$ group actions on the virtual level. The PEPS wavefunction for the Hofstadter model in the Landau gauge, as shown in Fig.~\ref{fig:hofstadter}, is therefore given by
\begin{equation} \label{eq:ansatz}
    \ket{\Psi_\phi(A)} = \diagram{9} \;,
\end{equation}
i.e., we have added virtual group actions in a specific pattern that matches the pattern of the hopping phases in the Hamiltonian. This ensures that the flux tensors contribute a cumulative phase of $\phi$ around each plaquette, matching the enclosed magnetic flux.

Note that our PEPS ansatz in Eq.~\ref{eq:ansatz} can be naturally generalized by combining the same flux pattern on the virtual bonds with an enlarged unit cell of tensors $A(x,y)$ of size $l_x\times l_y$. This allows one to capture phases with spontaneously broken translation symmetry, where the unit cell dimensions are not necessarily commensurate with the magnetic unit cell.

\parL{Perturbative expansion}%
%
One way of further motivating this PEPS ansatz is by perturbation theory \cite{Vanderstraeten2017}. In the infinite-$U$ limit at unit filling, the ground state is the simple product state with one boson on every site. For the Hofstadter-Hubbard Hamiltonian [Eq.~\eqref{eq:hof_hub}], first-order perturbation theory in $t/U$ can be implemented by the operator
\begin{align}
    T = 1 &+  \frac{t}{U} \sum_{\braket{ij}} \left(\e^{i\phi_{ij}} b_i\dag b_j\pdag + \hc \right) \nonumber + \mathcal{O}\left(\frac{t^2}{U^2}\right) \;,
\end{align}
where the phase pattern $\phi_{ij}$ depends on the gauge. We observe that the nearest-neighbor operator with the Peierls phase factor can be decomposed using a flux tensor at the virtual level :
\begin{equation} \label{eq:perturbation}
    \e^{+i\phi} b_i\dag b_j\pdag + \e^{-i\phi} b_j\dag b_i\pdag = \diagram{10}
\end{equation}
with $P$ encoding the creation and annihilation operators
\begin{equation}
    \diagram{11} = b\dag , \qquad \diagram{12} = b\pdag \;.
\end{equation}
These tensors can be encoded in a tensor network operator, which gives rise to an extensive PEPS wavefunction \cite{Vanderstraeten2017} when applied to the product state (see App.~\ref{sec:pert} for the explicit entries up to second-order perturbation theory). As a result, the flux tensors in Eq.~\eqref{eq:perturbation} will appear on the virtual level of the resulting PEPS, with the pattern matching the gauge choice in the Hamiltonian. When choosing the Landau gauge $\vA=-Bx\, \vect{e}_y$ as in Fig.~\ref{fig:hofstadter}, we find precisely the pattern of flux tensors as in the PEPS ansatz that we introduced above.

\parL{Symmetries}%
%
Let us now infer the symmetry properties with respect to the magnetic translation operators [Eq.~\ref{eq:Tv}]. Due to the local $\U(1)$ symmetry of the PEPS tensors [Eq.~\eqref{eq:u1_peps}], any gauge transformation [Eq.~\eqref{eq:gauge_trf}] can immediately be dragged to the virtual level of the PEPS. The $\U(1)$ group elements will recombine with the flux tensors on the virtual level, giving rise to a magnetic PEPS with a different pattern of virtual flux tensors (see App.~\ref{sec:gauge} for an explicit example of gauge transformation). By choosing the gauge transformation corresponding to a translation of the system, one can check that this is equivalent to a translation of the PEPS itself. This then shows that the PEPS ansatz is invariant under the magnetic translation operators, i.e., under lattice translations combined with appropriate gauge transformations that restore the original pattern of Peierls phases, hence mapping the tensor network onto itself. A very similar argument holds for the rotational symmetry of the model, under the condition that the PEPS tensors be chosen to be rotationally symmetric. We conclude that, although the PEPS wavefunction is itself not translation invariant because of the flux tensors, it does give rise to physical observables that respect all spatial symmetries of the model.

\parL{PEPS algorithm}%
%
Because the flux tensors break the translation symmetry, one expects that contracting this PEPS and computing its expectation values would still require large unit cell contraction algorithms. Here we now show that this is not the case: the contraction problem can be solved by a simple one-site boundary MPS algorithm. We focus on the row-to-row transfer matrix, which is given by
\begin{equation}
    \mathcal{T}_{\textmath{row}} = \diagram{13}
\end{equation}
with $a$ the double-layer tensor and the flux tensors now acting on two legs simultaneously. The boundary MPS is found as an approximation of the leading eigenvector or fixed point $\ket{\Phi}_{\textmath{fp}}$ of this operator \cite{Haegeman2017, Fishman2018, Vanderstraeten2022}, i.e. it approximates the fixed-point equation
\begin{equation} \label{eq:fp}
    \mathcal{T}_{\textmath{row}} \ket{\Phi}_{\textmath{fp}} \propto \ket{\Phi}_{\textmath{fp}} \;.
\end{equation}
In this equation, the layer of flux tensors is the only part that breaks the translational symmetry. We can observe, however, that the action of this flux layer on a boundary MPS does not break the translation symmetry:
\begin{multline}
    \diagram{14} \\= \diagram{15} \;,
\end{multline}
because of the $\U(1)$ symmetry of the boundary MPS tensors. This shows that the fixed-point equation \eqref{eq:fp} for the boundary MPS becomes entirely translation invariant :
\begin{multline}
    \diagram{16} \\ \propto \diagram{17} \;.
\end{multline}
This fixed-point equation differs from the usual one solely by the presence of the flux tensors on the MPS virtual level on the left-hand side. As a result, we can straightforwardly modify efficient numerical algorithms for finding the fixed-point tensor $M$, simply by adding virtual flux tensors into the equations. Here, the flux per plaquette $\phi$ only appears as a number in the contractions, and the dependence on the particular gauge choice has disappeared from the problem.

This shows that the variational energy of the PEPS can be evaluated by a single-site contraction algorithm that only depends on the PEPS tensor $A$ and the flux $\phi$. As a result, the energy gradient is also entirely translation invariant and can be evaluated by an explicit summation scheme \cite{Vanderstraeten2016, Vanderstraeten2022} or by performing backtracking using automatic differentiation \cite{Liao2019}. As a result, the variational energy can be minimized by a gradient-based optimization routine without breaking any spatial symmetries. This, finally, constitutes a variational PEPS algorithm that does not require choosing a particular gauge for the vector potential, because the magnetic flux only enters by adding flux tensors in a few instances of tensor contractions.

\parL{Benchmark}%
%
We showcase the performance of this single-site PEPS code by simulating the bosonic version of the square-lattice Harper-Hofstadter-Hubbard model [Eq.~\eqref{eq:hof_hub}] at unit filling and with interaction strength $U/t=20$ where we tune the magnetic flux per plaquette continuously in the range $0\to\pi$ and we truncate the maximal local occupation at two. The phase diagram of the model at unit filling is rather well known. At $\phi=0$ the model exhibits an insulating phase (large $U/t$) and a superfluid phase (small $U/t$) with a continuous phase transition around $(U/t)_c\approx16.7$ \cite{Capogrosso2008}. By turning on the magnetic field, the transition is pushed to smaller $U/t$, i.e., the field stabilizes the insulating phase without changing its qualitative features \cite{Umucalilar2007, Natu2016, Song2019}. To address the Mott phase, we use the ansatz in Eq.~\ref{eq:ansatz}, which enforces both $\U(1)$ and full translation symmetry. Phases that break translation symmetry \cite{Umucalilar2007, Goldbaum2008, Powell2011, Moeller2010, Natu2016} can be described within the same framework by employing enlarged unit cells, as discussed above.

\begin{figure}[t]
    \centering
    \includegraphics[width=0.9999\linewidth]{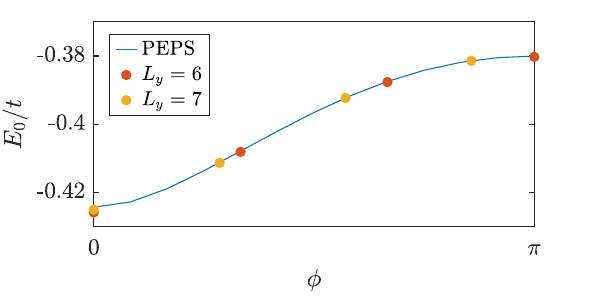}
    \caption{Ground state energy per site of the bosonic Harper-Hofstadter-Hubbard model [Eq.\eqref{eq:hof_hub}] at $U/t=20$ and unit filling as a function of the flux per plaquette $\phi$, computed with our variational PEPS ansatz on the infinite plane, compared to MPS simulation on an infinite cylinder of width $L_y=6$ and $L_y=7$.}
    \label{fig:bench}
\end{figure}

We perform variational optimization \cite{Vanderstraeten2016, Corboz2016} of a PEPS with bond dimension $D=4$. We compare with MPS simulations on the infinite cylinder with circumference $L_y$; in order to keep the MPS unit cell manageable, we are limited to discrete values of the flux $\phi=2\pi k/L_y$. We use variational optimization methods for MPS on the infinite cylinder \cite{ZaunerStauber2018}. The results in Fig.~\ref{fig:bench} show a very good agreement, confirming that our PEPS ansatz indeed captures correctly the physics of interacting bosons in a magnetic field.

\parL{Conclusions}%
%
In this work, we have formulated a PEPS ansatz for capturing ground states of interacting many-body lattice systems in a uniform magnetic field. The PEPS wavefunction is invariant under the action of the magnetic translation operators, and all observables are translation invariant by construction. Based on this ansatz, we construct a variational PEPS algorithm that preserves this translation symmetry, where the flux per plaquette enters as a continuous parameter. The practical utility of this method is manifest because we circumvent the need for large unit cells, and we can effectively tune the flux per plaquette as a continuous parameter without any complications. On a conceptual level, our tensor network approach allows us to study interacting many-body lattice systems without choosing a particular gauge for the vector potential. This fits into the larger program of constructing and optimizing the most efficient tensor network representations that combine all spatial and internal symmetries of a given phase \cite{Jiang2015, Mambrini2016, Hackenbroich2018, Hasik2024}. Note that we have specialized to interacting bosons for sake of simplicity, but the PEPS ansatz can be easily adapted to study interacting fermions in a uniform magnetic field by using fermionic tensor networks \cite{Mortier2025}.

A next step consists in considering non-integer fillings that potentially stabilize fractional Chern insulator (FCI) phases with chiral topological order. Formally, the PEPS ansatz [Eq.~\eqref{eq:ansatz}] can be generalized by imposing a fractional charge $p/q$ on the auxiliary bond, which necessitates also a fractionalization of the virtual charges and leads to the PEPS exhibiting a virtual $\mathds{Z}_q$ symmetry. Our preliminary simulations of these fractionally-filled PEPS indicate that these states can be contracted, but require large bond dimensions for attaining adequate precision on local expectation values -- this is already the case for a state without flux. Nonetheless, we expect that the fractionalized version of our PEPS ansatz in a magnetic field will be extremely useful for simulating FCI phases with PEPS, and shed new light on the ongoing debate on the possibility of representing states with chiral topological order with tensor networks \cite{Wahl2013, Dubail2015, Yang2015, Poilblanc2015, Chen2018, Hasik2022, Weerda2024}.

\parL{Acknowledgments}%
%
The authors would like to thank Nick Bultinck, Nathan Goldman, Felix Palm and Matteo Rizzi for interesting discussions. We acknowledge the EPSRC International Tensor Network for catalyzing multiple encounters that have enhanced this project.  W.T. is supported by Research Foundation Flanders (FWO) [Postdoctoral Fellowship 12AA225N]. GM acknowledges support by the Royal Society under University Research Fellowship award URF\textbackslash R\textbackslash 180004 and from Kent-Ghent project funding by the University of Kent. FV acknowledges funding from the UKRI grant EP/Z003342/1, BOFGOA (Grant No. BOF23/GOA/021), EOS (Grant No.~40007526), and IBOF (Grant No.~IBOF23/064).

\bibliography{bibliography.bib}

\appendix

\section{Perturbative expansion}
\label{sec:pert}

\renewcommand{\diagram}[1]{\vcenter{\hbox{\includegraphics[scale=0.6,page=#1]{./diagrams_perturbation.pdf}}}}

A general procedure for finding the most natural structure for a tensor network representation in a given gapped quantum phase is by perturbation theory \cite{Vanderstraeten2017}: Starting from a point in the phase with a well-known tensor network representation (typically, an RG fixed point), one can write down a perturbative expansion in terms of an extensive tensor network operator. When applied onto the starting point, the result is a tensor network state with a certain (spatial and internal) symmetry structure. The structure of this tensor network representation can then be used throughout the whole phase, but with enlarged variational spaces. 

For the current problem of the Harper-Hofstadter-Hubbard model in Eq.~\eqref{eq:hof_hub}, we start from the infinite-$U$ limit at unit filling, where the ground state is the simple product state with one boson on every site (regardless of the flux per plaquette) : 
\begin{equation}
    \ket{\Psi_0} = \diagram{1} \;,
\end{equation}
where the one-site tensors is a simple local product state
\begin{equation}
    \diagram{2} = \delta_{p,q_a} \;,
\end{equation}
and the dashed lines are trivial charge-zero lines. The first-order perturbation theory in $t/U$ can be implemented by the operator
\begin{align}
    T = 1 &+  \frac{t}{U} \sum_{\braket{ij}} \left(\e^{i\phi_{ij}} b_i\dag b_j\pdag + \hc \right) \nonumber \\
    &+ \frac{t^2}{2U^2} \sum_{\braket{ij}\braket{kl}} \left(\e^{i\phi_{ij}} b_i\dag b_j\pdag + \hc \right)\left(\e^{i\phi_{kl}} b_k\dag b_l\pdag + \hc \right) \nonumber \\
    &+ \frac{t^2}{2U^2} \sum_{\braket{ijk}} \left(\e^{i\phi_{ij}} b_i\dag b_j\pdag + \hc \right)\left(\e^{i\phi_{jk}} b_j\dag b_k\pdag + \hc \right) \nonumber \\
    &+ \mathcal{O}(t^3/U^3),
\end{align}
where the second sum runs over non-overlapping nearest-neighbor pairs, and the third sum runs over three-site clusters. The nearest-neighbor operator is represented as a tensor that can be decomposed as:
\begin{equation}
    \e^{+i\phi} b_i\dag b_j\pdag + \e^{-i\phi} b_j\dag b_i\pdag = \diagram{3}
\end{equation}
with $P$ encoding the creation and annihilation operators
\begin{equation}
    \diagram{4} = b\dag , \qquad \diagram{5} = b\pdag \;.
\end{equation}
In the tensor network encoding of the phase-assisted hopping, we observe the $\U(1)$ group action appearing on the virtual level. In second order in perturbation theory, there will appear three-site clusters represented by a tensor contraction of the form
\begin{align}
    \left(\e^{\phi_{\braket{ij}}} b_i\dag b_j\pdag\right) & \left(\e^{\phi_{\braket{jk}}} b_j\dag b_k\pdag\right) \nonumber \\
    &= \e^{i(\phi_{\braket{ij}}+\phi_{\braket{jk}})} b_i\dag n_j b_k\pdag \nonumber  \\
    &= \diagram{6} \;,
\end{align}
where we recycle the $P$ tensors from the first-order term, and we introduce a new tensor $Q$ containing the local number operator $n$,
\begin{equation}
    \diagram{7} = \diagram{8} = n \;.
\end{equation}
The first- and second-order perturbation theory can therefore be represented as an extensive tensor network operator \cite{Vanderstraeten2017} of the form
\begin{equation} \label{eq:tno}
    T = \diagram{9} \;,
\end{equation}
with the entries in Tab.~\ref{tab:tno}. When applying this operator to the Mott-insulator product state at a certain filling encoded by the auxiliary legs, we arrive at the PEPS ansatz [Eq.~\eqref{eq:ansatz}]. From this perturbative construction, it can be readily seen that the ansatz captures the presence of the magnetic field in the Landau gauge, but without breaking any symmetries of the model.

\renewcommand{\diagram}[1]{\vcenter{\hbox{\includegraphics[scale=0.5,page=#1]{./diagrams_perturbation.pdf}}}\,}

\begin{table}[t]
    \centering
    \begin{tabular}{|l|l|}
    \hline
    \multicolumn{2}{|l|}{$\;\diagram{10} =1 $} \\
    \hline
    $\;\diagram{11} = \cdots = \lambda_1 b\dag \;$  &  $\;\diagram{12} = \cdots = \lambda_1 b\pdag \;$   \\
    \hline
    \multicolumn{2}{|l|}{$\;\diagram{13}  = \diagram{14}  = \cdots = \lambda_2 n $} \\
    \hline
    \end{tabular}
    \caption{The entries in the tensor network operator representing first- and second-order perturbation theory, where $\lambda_1\propto t/U$ and $\lambda_2\propto (t/U)^2$; the dotes denote permutations of the leg labels getting the same entry. Upon insertion of these entries in the operator in Eq.~\ref{eq:tno}, we get a perturbative expansion of the Hofstadter-Hubbard model in Eq.~\ref{eq:hof_hub}.}
    \label{tab:tno}
\end{table}

\section{Gauge transformation}
\label{sec:gauge}

\renewcommand{\diagram}[1]{\vcenter{\hbox{\includegraphics[scale=0.6,page=#1]{./diagrams_gauge.pdf}}}}

In this appendix, we show explicitly that a gauge transformation of the form \eqref{eq:gauge_trf_U} leads to a reconfiguration of the virtual flux tensors in the PEPS ansatz by using the pull-through condition [Eq.~\eqref{eq:u1_peps}] and charge-conserving bond tensors [Eq.~\eqref{eq:round}]. 
We start from the PEPS ansatz in Eq.~\eqref{eq:ansatz},
\begin{align}
    \ket{\Psi(A)} = \diagram{1},
\end{align}
representing a wavefunction for the ground state of the Harper-Hofstadter-Hubbard Hamiltonian in the Landau gauge $\vA=-Bx\, \vect{e}_y$ (i.e. the Peierls phase pattern indicated in Fig.~\ref{fig:hofstadter}). At a certain row in the lattice, we start absorbing the virtual flux tensors. For every absorption, we use the $\U(1)$ symmetry of the PEPS tensor as in Eq.~\eqref{eq:u1_peps}. After absorption, we find the following pattern of flux tensors :
\begin{equation}
    \ket{\Psi(A)} \propto \diagram{2}.
\end{equation}
Here we have used the fact that a flux tensors changes sign when crossing the round tensor (which is defined in Eq.~\eqref{eq:round}), and recombines with the flux tensor on the other side:
\begin{equation}
    \diagram{5} \propto \diagram{6} \;.
\end{equation}
We observe that the absorption of the virtual flux tensors also has an effect on the physical degrees of freedom. Note that we have not explicitly indicated the action of the $\U(1)$ group actions on the auxiliary legs (i.e., the wavy legs), because this just yields a global phase factor to the wavefunction which does not play a role for a wavefunction in the thermodynamic limit. 

Absorbing a next row of tensors yields
\begin{equation}
    \ket{\Psi(A)} = \diagram{3},
\end{equation}
where we have again used the recombination of flux tensors on either sides of the round tensors on the horizontal bonds and have omitted the global phase factors from the $\U(1)$ group actions on the wave legs. Finally, we absorb a next row of flux tensors:
\begin{equation}
    \ket{\Psi(A)} = \diagram{4}.
\end{equation}
We see a clear pattern emerging: sequentially absorbing the flux tensors on the vertical auxiliary legs (which corresponded to the pattern of Peierls phases in the Landau gauge $\vA=-Bx\, \vect{e}_y$), gives rise to a new pattern of flux tensors, now on the horizontal virtual legs, corresponding to the pattern of Peierls phases in the other Landau gauge $\vA=By\vect{e}_x$. The pattern of $\U(1)$ group actions on the physical degrees of freedom is precisely the unitary gauge transformation in Eq.~\eqref{eq:gauge_trf_U} that connects the two different Landau gauges.

\end{document}